\documentstyle[prb,aps,preprint]{revtex}
\begin{document}
\draft
\title{BCS and Generalized BCS Superconductivity in Relativistic Quantum Field Theory. I. Formulation \footnote{This work is a part of the doctoral thesis of the author presented to Osaka university, Nov. 2000.}}
\author{Tadafumi Ohsaku}
\address{Department of Physics, Graduate School of Science, Osaka University, Machikaneyama-cho 1-1, Toyonaka, Osaka, 560-0043, Japan}
\maketitle

\newcommand{\bmx}{\mbox{\boldmath $x$}}
\newcommand{\bmy}{\mbox{\boldmath $y$}}
\newcommand{\bmk}{\mbox{\boldmath $k$}}
\newcommand{\bmp}{\mbox{\boldmath $p$}}
\newcommand{\bmq}{\mbox{\boldmath $q$}}
\newcommand{\bmP}{\mbox{\boldmath $P$}}  
\newcommand{\kfey}{\ooalign{\hfil/\hfil\crcr$k$}}
\newcommand{\pfey}{\ooalign{\hfil/\hfil\crcr$p$}}
\newcommand{\qfey}{\ooalign{\hfil/\hfil\crcr$q$}}
\newcommand{\Deltafey}{\ooalign{\hfil/\hfil\crcr$\Delta$}}  
\def\sech{\mathop{\rm sech}\nolimits}

\begin{abstract}

We investigate the BCS and generalized BCS theories in the relativistic quantum field theory. We select the gauge freedom as $U(1)$, and introduce a BCS-type effective attractive interaction. After introducing the Gor'kov formalism and performing the group theoretical consideration of the mean fields, we solve the relativistic Gor'kov equation and obtain the Green's functions in analytical forms. We obtain various types of gap equations.

\end{abstract}

\section{Introduction}
\label{sec:intro}

The most important turning point in the history of the theory of superconductivity is the appearance of the BCS ( Bardeen-Cooper-Schriefer ) theory~[1]. It is in fact the first success giving an explanation of the superconductivity from the microscopic dynamics of the interacting many-fermion system. From a view point of the condensed matter theory, the BCS theory can be interpreted as a simplification of the electron-phonon theory, which reveals the fundamental origin of the mechanism of the generation of the superconducting state. However, the influence of the BCS theory is much more wide. It is the famous fact that its influence reached not only to the condensed matter physics, but also to the elementary particle physics and the nuclear physics~[2$\sim$6]. 

The BCS theory succeeded in explaining the superconductivity in metals and alloys. But in the development of the high-$T_{c}$ and heavy-fermion superconductivities, the BCS theory could not cover them completely. This situation caused new discussions on the mechanism of the origin of these superconductivities~[7]. Apart from these discussions on the mechanism of the origin, various efforts also have been made to extend ( generalized ) the BCS theory to understand these systems~[8$\sim$15]. In the generalized BCS theory, the question of the mechanism of the origin is set aside for a while, and the emphasis is placed on the symmetries of the pairing interaction and the mean fields. They extended the treatment of the Cooper pairs, not only to the singlet pairing but also to the triplet pairing, in which the mean fields have the vectorial nature; they have finite angular momenta. 
The difference of the attitudes between the theories of the mechanisms of superconductivities and the generalized BCS approach has an important meaning in the theory of superconductivity.

Until now, almost all the theories of condensed matter physics are constructed on the nonrelativistic quantum theory. On the other hand, in the electronic structure calculations based on the band theory, it is obvious that, in the cases of heavy elements, we have to take into account several relativistic effects for correct descriptions even qualitatively. For example, even in the case of slow electrons like ones with a kinetic energy of 100 eV ( the electron mass is about 0.511 MeV ), we need the 4-component Dirac equation for an qualitatively correct description of the system~[16]. Needless to say, the targets of condensed matter theory are more or less to include the relativistic effect. However, the study on the relativistic effects in condensed matter is not so much performed, except the considerations of the spin-orbit coupling~[17]. 

Under this situation, recently papers which assert the necessity of the relativistic treatment for superconductivity in condensed matter were published by K. Capelle and E. K. U. Gross~[18$\sim$24].  The heavy elements like ${\rm Au}$, ${\rm Pb}$, ${\rm Ir}$ or ${\rm I}$ become superconducting states~[25], and simultaneously they are under strong relativistic effects. The high-$T_{c}$ or heavy-fermion superconductors normally contain heavy elements like ${\rm La}$, ${\rm Ce}$ and ${\rm U}$. Especially in the heavy-fermion superconductors, the f-electrons which are under strong relativistic effects, become the superconducting state. In principle, the description of the electronic structure for these systems requires the relativistic treatment. It is also a well known fact that, the spin-orbit effect plays a role in the NMR spectra ( the Knight shift ) and also in the impurity effects in superconductors~[17]. Naturally it becomes interesting to treat these physics with the strong relativistic effects, beyond making a small corrections. There are no reasons that those systems which have strong relativistic effects do not show relativistically distinct superconductivity than the nonrelativistic one. In the works of Capelle and Gross~[18,19], the necessity of the relativistic treatment of the theory of superconductivity in solid state physics was argued in the same context given above. They introduced the Dirac type Bogoliubov-de Gennes Hamiltonian and discussed the group theoretical aspects of the Cooper pair, the mean fields. They also discussed the response of superconductors to circularly polarized light~[22]. P. Strange discussed the gap equation based on the works of Capelle-Gross~[26]. Therefore, we wish to extend their theory to a more general form, and discuss various aspects of the relativistic superconductivity. 

Recently, the relativistic superfluid theory is developed in quantum hadrodynamics ( the theory for the system of protons, neutrons and mesons )~[27,28]. Also recently, in the quark matter in quantum chromodynamics ( QCD ), the possibility of the formation of the Cooper pair ( the color-superconductivity ) is discussed~[29,30]. These theories treat the superconductivity and/or superfluidity in relativistic many-body systems. The neutron stars~[31] may become a subject of these theories. The superconductivity in the high-density plasma should also be treated in relativistic theory, though we do not find any of this attempt in literature.   

With these tendencies of recent theoretical physics, and considerations given above, we investigate the BCS and generalized BCS theories in relativistic quantum field theory in this paper. Needless to say, the most important approach in the theory of superconductivity is the BCS and generalized BCS schemes. The essential idea of the BCS theory is constructed on an effective model Lagrangian, like the Nambu-Jona-Lasinio model or the Gross-Neveu model~[6]. In our work, we introduce the relativistic BCS Lagrangian, and study the dynamical $U(1)$-gauge symmetry breaking. The attractive interaction should arise from some physical processes in a matter, as discussed in the theory of the mechanism of the origin of the superconductivity. Here we obey the attitude of the generalized BCS theory, and construct the BCS and generalized BCS theory in the relativistic framework for studying various systems.  Since we select the gauge freedom as $U(1)$, our main subject is many-electron-positron systems. We try to keep a close relation with condensed matter theory. Namely, we select the simplest formalism and intend to study a general theory for the relativistic superconductivity. But our formalism can also be applied to the color-superconductivity in QCD. In this work, we are mainly interested in the methodological aspect. We want to present how the relativistic theory of the superconductivity is constructed and what characteristics it has. {\it We intend to construct an unified approach about the superconductivity in the systems under strong relativistic effects, from solid state physics to the nuclear and particle physics.} Due to the standing on the attitude of the BCS and generalized BCS theory, we can realize this purpose. 

This paper is organized as follows. In Sec. \ref{sec:for}, the Gor'kov formalism~[32] is introduced from a Lagragian as a starting point. We regard that this Lagrangian gives nonrelativistic BCS theory in its nonrelativistic limit. In Sec. \ref{sec:gro}, we perform the group theoretical consideration of the mean fields in complete manner. In Sec. \ref{sec:gen}, especially for treating the spin triplet pairing, the generalized BCS theory is constructed on the basis of the previous results. In Sec. \ref{sec:sol}, we solve our Gor'kov equations which have several different types of the mean fields, and obtain the Green's functions in analytical forms. In Sec. \ref{sec:gap}, we give the gap equations for various states and discuss their features. Finally in Sec. \ref{sec:con}, we give summary of this work, and also give a plan about coming works in near future. The numerical part will be published as part II of this study.

\section{Relativistic Gor'kov Equation}
\label{sec:for}

The purpose of this section is to introduce the basis of the theory of the relativistic BCS superconductivity by using the field theoretical Green's function method. For this procedure, following the nonrelativistic theory, we use the canonical formalism. In particular, we select the Gor'kov formalism~[32], because it is the most fundamental formalism in superconductivity, and also because we want to obtain the Green's function which increases the ability of the applications. Another purpose is to construct the theory which will be refered to in the generalized BCS formalism. 

First, we introduce the following Lagrangian as our starting point:
\begin{eqnarray}
{\cal L}(x) &=& -\frac{1}{4}F_{\mu\nu}(x)F^{\mu\nu}(x) + \bar{\psi}(x)i\gamma^{\mu}(\partial_{\mu}-ieA_{\mu}(x))\psi(x)-m\bar{\psi}(x)\psi(x) \nonumber \\
& & +\frac{g_{0}}{2}(\bar{\psi}(x)\psi(x))^{2}, 
\end{eqnarray}
where the first term is the kinetic term of the electromagnetic field, $F_{\mu\nu}$ is the antisymmetric electromagnetic tensor, and $A_{\mu}$ is the electromagnetic potential. The second term is the kinetic term of the Dirac field, the third is the Dirac mass term, $\psi$ and $\bar{\psi}$ are the 4-component bispinors describing the Dirac fields. We select $\gamma^{0\dagger}=\gamma^{0}$, $\gamma^{i\dagger}=-\gamma^{i}(i=1,2,3)$. We use the metric convention as $g_{\mu\nu}={\rm diag}(1,-1,-1,-1)$ and hence $\gamma^{\mu}\gamma^{\nu}+\gamma^{\nu}\gamma^{\mu}=2g^{\mu\nu}$. The main object of our theory is electron-positron gas. Since generally we treat the system in which the numbers of electrons and positrons are different ( finite density ), we have to add the next term to our Lagrangian, 
\begin{eqnarray}
J^{\mu}(x)A_{\mu}(x), 
\end{eqnarray}
as the fifth term. This term is the background to neutralize and stabilize the system. $J^{\mu}$ is a classical current. This term, however, is not used to describe the physics hereafter. The fourth term of the Lagrangian is the four-body contact interaction at the same space-time point. It will describe attractive interaction between fermions with the coupling constant $g_{0}>0$. This term can be called the relativistic BCS interaction, and we take here the simplest form. We regard this Lagrangian involves the usual BCS theory in its nonrelativistic limit. To make it clear, we here take the standard representation:
\begin{equation}
\psi = \left(
\begin{array}{c}
\phi \\
\chi 
\end{array}
\right).
\end{equation}
Here, $\phi$ is the large component while $\chi$ is the small component. 
The small component becomes negligible at the nonrelativistic limit, and therefore the interaction $\frac{g_{0}}{2}(\bar{\psi}\psi)^{2}$ becomes $\frac{g_{0}}{2}(\phi^{\dagger}\phi)^{2}$. 
From the view point of the condensed matter physics, this Lagrangian is not completly microscopic, such as the electron-phonon theory, but should be regarded as a phenomenological effective Lagrangian. However, if once we introduce this Lagrangian, the dynamics of the system is completely determined from it. This theory is, of course, unrenormalizable, and we have to introduce a cut-off. This Lagrangian itself has symmetries of Poincar\'{e} invariance, $U(1)$-gauge invariance, charge-conjugation invariance, spatial inversion  and time-reversal invariance. 

Now we obtain the field equations from the Lagrangian by the action principle: \begin{eqnarray}
0 &=& \frac{\partial{\cal L}}{\partial \bar{\psi}}-\partial_{\mu}\frac{\partial{\cal L}}{\partial(\partial_{\mu} \bar{\psi})} \nonumber \\
&=& i\gamma^{\mu}(\partial_{\mu}-ieA_{\mu})\psi-m\psi+g_{0}(\bar{\psi}\psi)\psi, \\
0 &=& \frac{\partial {\cal L}}{\partial \psi}-\partial_{\mu}\frac{\partial {\cal L}}{\partial (\partial_{\mu}\psi)} \nonumber \\
&=& -i(\partial_{\mu}+ieA_{\mu})\bar{\psi}\gamma^{\mu}-m\bar{\psi}+g_{0}(\bar{\psi}\psi)\bar{\psi},
\end{eqnarray}			
and the Hamiltonian becomes 
\begin{eqnarray}
H &=& \int d^{3}\bmx\bar{\psi}(-i\gamma\cdot\nabla+m-e\gamma^{\mu}A_{\mu}-\gamma^{0}\mu)\psi -\frac{g_{0}}{2}\int d^{3}\mbox{\boldmath $x$}(\bar{\psi}\psi)^{2}. 
\end{eqnarray}
Here the second term will give the attractive interaction for $g_{0}>0$. Because we consider a many-body problem, we introduce the chemical potential $\mu$ in the above Hamiltonian. It descibes the finite density at $\mu\ne0$ as the conjugate of the particle number minus antiparticle number. Throughout this study, we treat $\mu$ as a parameter introduced from the outside of the system. Thus we set $\mu =\epsilon_{F}$ ( Fermi level ). Now, to obtain the Gor'kov equation, we use the equations of motion ( Eqs. (4) and (5) ) of the following form:
\begin{eqnarray}
i\gamma^{\mu}(\partial_{\mu}-ieA_{\mu})\psi-m\psi+g_{0}(\bar{\psi}\psi)\psi+\mu\gamma^{0}\psi &=& 0, \\
-i(\gamma^{\mu})^{T}(\partial_{\mu}+ieA_{\mu})\bar{\psi}^{T}-m\bar{\psi}^{T}+g_{0}(\bar{\psi}\psi)\bar{\psi}^{T}+\mu(\gamma^{0})^{T}\bar{\psi}^{T} &=& 0,
\end{eqnarray}
where $T$ means the transposition, and the second equation is obtained from (5) by taking the transposition.  

Now we employ the canonical quantization about the Dirac fields to switch to the quantized fields. Namely, we demand the anticommutation relations 
\begin{eqnarray}
\{\hat{\psi}_{\alpha}(x_{0},\mbox{\boldmath $x$}),\hat{\psi}_{\beta}^{\dagger}(x_{0},\mbox{\boldmath $y$})\} &=& \delta_{\alpha\beta}\delta^{(3)}(\mbox{\boldmath $x$}-\mbox{\boldmath $y$}), \\
\{\hat{\psi}_{\alpha}(x_{0},\bmx),\hat{\psi}_{\beta}(x_{0},\bmy)\} &=& \{\hat{\psi}^{\dagger}_{\alpha}(x_{0},\bmx),\hat{\psi}^{\dagger}_{\beta}(x_{0},\bmy)\} = 0. 
\end{eqnarray}
Here $\alpha$ and $\beta$ are the spinor indices. We do not quantize $A_{\mu}$ but treat it as a classical field. 

Next, we make some preparations to study the theory of superconductivity. First, we introduce various propagators. We use the 8-component Nambu notation~[33]: 
\begin{eqnarray}
 \hat{\Psi}(x) \equiv \left(
 \begin{array}{c}
 \hat{\psi}(x) \\
\hat{\bar{\psi}}^{T}(x)
\end{array}
\right), \quad \hat{\bar{\Psi}}(x) &\equiv& (\hat{\bar{\psi}}(x),\hat{\psi}^{T}(x)).   
\end{eqnarray}
Here $\psi$ and $\bar{\psi}$ are in the Heisenberg representation. The definition of the generalized one-particle propagator is
\[
{\bf G}(x,y) \equiv -i\langle T\hat{\Psi}(x)\hat{\bar{\Psi}}(y) \rangle
\]
\begin{equation} = \left(
 \begin{array}{cc}
 -i\langle T\hat{\psi}_{\alpha}(x)\hat{\bar{\psi}}_{\beta}(y)\rangle & -i\langle T\hat{\psi}_{\alpha}(x)\hat{\psi}^{T}_{\beta}(y)\rangle \\
 -i\langle T\hat{\bar{\psi}}^{T}_{\alpha}(x)\hat{\bar{\psi}}_{\beta}(y)\rangle & -i\langle T\hat{\bar{\psi}}^{T}_{\alpha}(x)\hat{\psi}^{T}_{\beta}(y)\rangle 
\end{array}
\right) 
= \left(
 \begin{array}{cc}
 S_{F}(x,y)_{\alpha\beta} & -iF(x,y)_{\alpha\beta}  \\
 -i\bar{F}(x,y)_{\alpha\beta} & -S_{F}(y,x)_{\beta\alpha}
\end{array}
\right). 
\end{equation}
This is a 8$\times$8 matrix. $T$ means the time-ordered product, and $\langle \cdots\rangle$ means the sum over the expectation value between states coupled through the superconducting pair-correlation. $S_{F}$ is the Feynman propagator for quasiparticle, $-iF$ and $-i\bar{F}$ are the anomalous propagators. Next, we obtain the equations of motion for the propagators (12), in a similar way as the nonrelativistic case. We employ the Gor'kov factorization in (7) and (8), taking account of only the superconducting pair-correlation by introducing the mean-field approximation. Then we obtain the relativistically generalized Gor'kov equation written down as an 8$\times$8 matrix equation, ${\bf L}(x){\bf G}^{Gor'kov}(x,y)=\hat{1}\delta^{(4)}(x-y)$, namely, 
\[
\left(
\begin{array}{cccc}
i\gamma^{\mu}(\partial_{\mu}-ieA_{\mu})-m+\gamma^{0}\mu & \Delta(x) \\
\bar{\Delta}(x) & i\gamma^{\mu T}(\partial_{\mu}+ieA_{\mu})+m-\gamma^{0T}\mu
\end{array}
\right)
\left(
\begin{array}{cccc}
S_{F}(x,y) & -iF(x,y) \\
-i\bar{F}(x,y) & -S_{F}(y,x)^{T}
\end{array}
\right)
\]
\begin{equation} 
=
\left(
\begin{array}{cccc}
\delta^{(4)}(x-y) & 0 \\
0 & \delta^{(4)}(x-y)
\end{array}
\right).
\end{equation} 
$\Delta(x)$ and $\bar{\Delta}(x)$ are 4$\times$4 matrix mean fields, so called order parameters. The definitions are
\begin{eqnarray}
\Delta(x_{0},\mbox{\boldmath $x$})_{\alpha\beta} &\equiv& g_{0}F(x^{+}_{0},\mbox{\boldmath $x$};x_{0},\mbox{\boldmath $x$})_{\alpha\beta}  = g_{0}\langle\hat{\psi}_{\alpha}(x^{+}_{0},\bmx)\hat{\psi}^{T}_{\beta}(x_{0},\bmx)\rangle,   \\
\bar{\Delta}(x_{0},\bmx)_{\alpha\beta} &\equiv& g_{0}\bar{F}(x^{+}_{0},\bmx;x_{0},\bmx)_{\alpha\beta} = g_{0}\langle \hat{\bar{\psi}}^{T}_{\alpha}(x^{+}_{0},\bmx)\hat{\bar{\psi}}_{\beta}(x_{0},\bmx)\rangle.
\end{eqnarray}
This gives the self-consistency conditions. Eq. (13) is determined self-consistently. Needless to say, we can obtain infinite order series of the Dyson-type equation from (13).
In the superconducting state ( or under the critical temperature ), the mean fields have finite values, and we stand on the theory which does not conserve the particle number and charge; we attain different representation by non-perturbative way. In general, the mean field clearly violates the Lorentz symmetry, as well as the gauge symmetry. In other words, the mean field involves quantities other than the scalar. In the case of ``superconductivity'' theory in particle physics, the object is the vacuum: They treat only the scalar value $\langle\bar{\psi}\psi\rangle\ne 0$ ( the Hartree fields ) and they discuss the chiral symmetry breaking~[2,5]. These points are different from our theory. We would like to mention the relation between our Gor'kov equation (13) and the relativistic Bogoliubov-de Gennes equation which was given in ref.[18]. If we explicitly write the Coulomb potential of nuclei for $A_{\mu}$ of (13), our theory becomes essentially the same as the relativistic Bogoliubov-de Gennes theory.

We will also obtain the Fourier transform of the Gor'kov equation for the homogeneous system. We set the external field $A_{\mu}=0$, and then obtain it in the matrix form:
\begin{equation} \left(
\begin{array}{cccc}
\kfey-m+\gamma^{0}\mu & \Delta \\
\bar{\Delta} & \kfey^{T} +m-\gamma^{0T}\mu 
\end{array}
\right)
\left(
\begin{array}{cccc}
S_{F}(k) & -iF(k) \\
-i\bar{F}(k) & -S_{F}(-k)^{T}
\end{array} 
\right)
=
\left(
\begin{array}{cccc}
1 & 0 \\
0 & 1 
\end{array}
\right).
\end{equation}
Here $\kfey^{T}$ means the transpose of $\kfey$. The self-consistency condition now becomes
\begin{eqnarray}
\Delta &=& g_{0}\int\frac{d^{4}p}{(2\pi)^{4}}F(p), \\
\bar{\Delta} &=& g_{0}\int\frac{d^{4}p}{(2\pi)^{4}}\bar{F}(p).
\end{eqnarray}
Because of the homogeneity, the mean field has only the internal degrees of freedom. In the nonrelativistic BCS theory, the mean field has no internal degree of freedom. In the case of the relativistic theory, there is a possibility to obtain much more complicated states. 

The finite-temperature theory of the Matsubara formalism can be obtained in the same way. We introduce imaginary time $\tau=it$. The temperature Green's functions are defined as
\begin{eqnarray}
{\bf {\cal G}}(x,y) &\equiv& -\langle T_{\tau}\hat{\Psi}(x)\hat{\bar{\Psi}}(y)\rangle \nonumber
\end{eqnarray}
\begin{eqnarray}
= 
\left(
\begin{array}{cccc}
-\langle T_{\tau}\hat{\psi}_{\alpha}(x)\hat{\bar{\psi}}_{\beta}(y)\rangle & -\langle T_{\tau}\hat{\psi}_{\alpha}(x)\hat{\psi}^{T}_{\beta}(y)\rangle \\
-\langle T_{\tau}\hat{\bar{\psi}}^{T}_{\alpha}(x)\hat{\bar{\psi}}(y)_{\beta} \rangle & -\langle T_{\tau}\hat{\bar{\psi}}^{T}_{\alpha}(x)\hat{\psi}^{T}_{\beta}(y)\rangle 
\end{array}
\right)
=
\left(
\begin{array}{cccc}
{\cal S}(x,y)_{\alpha\beta} & -{\cal F}(x,y)_{\alpha\beta} \\
-\bar{{\cal F}}(x,y)_{\alpha\beta} & -{\cal S}(y,x)_{\beta\alpha}
\end{array}
\right),
\end{eqnarray}
where $\langle \cdots\rangle$ means the statistical average. From the equations of motion of the temperature Green's functions, the Gor'kov equation becomes
\[
\left(
\begin{array}{cccc}
-\gamma^{0}(\frac{\partial}{\partial\tau}-\mu)+i\gamma^{k}\partial_{k}-m  & \Delta(x) \\
\bar{\Delta}(x) & -\gamma^{0T}(\frac{\partial}{\partial\tau}+\mu)+i\gamma^{kT}\partial_{k}+m 
\end{array}
\right)
\left(
\begin{array}{cccc}
{\cal S}(x,y)_{\alpha\beta} & -{\cal F}(x,y)_{\alpha\beta} \\
-{\bar {\cal F}}(x,y)_{\alpha\beta} & -{\cal S}(y,x)_{\beta\alpha}
\end{array}
\right)
\]
\begin{equation}
=
\left(
\begin{array}{cccc}
\delta^{(4)}(x-y) & 0 \\
0 & \delta^{(4)}(x-y) 
\end{array}
\right).
\end{equation}
Here the definition of the mean fields are the simple extension of those for the zero temperature:
\begin{eqnarray}
\Delta(\tau,\bmx)_{\alpha\beta} &\equiv& g_{0}{\cal F}(\tau^{+},\bmx;\tau,\bmx) = g_{0}\langle\hat{\psi}_{\alpha}(\tau^{+},\bmx)\hat{\psi}^{T}_{\beta}(\tau,\bmx)\rangle, \\
\bar{\Delta}(\tau,\bmx)_{\alpha\beta} &\equiv& g_{0}{\bar {\cal F}}(\tau^{+},\bmx;\tau,\bmx) = g_{0}\langle \hat{\bar{\psi}}^{T}_{\alpha}(\tau^{+},\bmx)\bar{\psi}_{\beta}(\tau,\bmx) \rangle. 
\end{eqnarray}
Fourier transform for the homogeneous system are also obtained, and the Gor'kov equation in the matrix form becomes
\[
\left(
\begin{array}{cccc}
\gamma^{0}(i\omega_{n}+\mu)-\gamma\cdot\bmk-m & \Delta \\
\bar{\Delta} & \gamma^{0T}(i\omega_{n}-\mu)-\gamma^{T}\cdot\bmk+m 
\end{array}
\right)
\left(
\begin{array}{cccc}
{\cal S}(\omega_{n},\bmk) & -{\cal F}(\omega_{n},\bmk) \\
-{\bar{\cal F}}(\omega_{n},\bmk) & -{\cal S}(-\omega_{n},-\bmk)^{T}
\end{array}
\right)
\]
\begin{equation}
=
\left(
\begin{array}{cccc}
1 & 0 \\
0 & 1
\end{array}
\right).
\end{equation}
Here $\beta\equiv1/k_{B}T$ ( $k_{B}$; the Boltzmann constant ), $\omega_{n}=(2n+1)\pi/\beta$ is a fermion discrete frequency. Solving the finite-temperature Gor'kov equation, we will obtain the solutions of the same form, except that we need to substitute $k_{0}\to i\omega_{n}$. The solutions of the Gor'kov equation for homogeneous system are discussed in Sec. V.

\section{Group theoretical consideration of the mean fields}
\label{sec:gro}

We proceed the group theoretical consideration so as to treat the mean fields more easily. From the anticommutation relation for Fermi fields, the mean fields obey
\begin{eqnarray}
\Delta(x)_{\alpha\beta} = -\Delta(x)_{\beta\alpha}, \quad \bar{\Delta}(x)_{\alpha\beta} = -\bar{\Delta}(x)_{\beta\alpha}.
\end{eqnarray}
Note that $\bar{\Delta}$ is not the Hermitian conjugate of $\Delta$:
\begin{eqnarray}
\bar{\Delta}(x) = g_{0}\langle \bar{\psi}^{T}(x)\bar{\psi}(x)\rangle = g_{0}\gamma^{0T}\langle\psi^{\dagger}(x)\psi^{\dagger}(x)\rangle\gamma^{0} = \gamma^{0}\Delta^{\dagger}(x)\gamma^{0}, 
\end{eqnarray}
but rather, $\gamma^{0}\bar{\Delta}$ is the Hermitian conjugate of $\gamma^{0}\Delta$. On the other hand, from the gauge transformation: 
\begin{eqnarray}
\psi(x) \to \psi'(x) = e^{i\alpha(x)}\psi(x), \quad \bar{\psi}(x) \to \bar{\psi}'(x) = \bar{\psi}(x)e^{-i\alpha(x)}, 
\end{eqnarray} 
the one-body propagator is transformed as
\[
\left(
\begin{array}{cccc}
S_{F}(x,y) & -iF(x,y) \\
-i\bar{F}(x,y) & -S_{F}(y,x)^{T} 
\end{array}
\right)
\to
\left(
\begin{array}{cccc}
e^{i(\alpha(x)-\alpha(y))}S_{F}(x,y) & e^{i(\alpha(x)+\alpha(y))}(-i)F(x,y) \\ 
e^{-i(\alpha(x)+\alpha(y))}(-i)\bar{F}(x,y) & e^{-i(\alpha(x)-\alpha(y))}(-)S_{F}(y,x)^{T}
\end{array}
\right),
\]
\begin{equation}
\end{equation}
and the mean fields are transformed as
\begin{eqnarray}
\Delta(x)_{\alpha\beta} \to e^{2i\alpha(x)}\Delta(x)_{\alpha\beta}, \quad \bar{\Delta}(x)_{\alpha\beta} \to e^{-2i\alpha(x)}\bar{\Delta}(x)_{\alpha\beta}.
\end{eqnarray}
Under the spatial inversion,
\begin{eqnarray}
\psi(x_{0},\bmx) \stackrel{\cal P}{\to} \gamma^{0}\psi(x_{0},-\bmx), \quad \bar{\psi}(x_{0},\bmx) \stackrel{\cal P}{\to} \bar{\psi}(x_{0},-\bmx)\gamma^{0}, 
\end{eqnarray}
the mean fields are transformed as
\begin{eqnarray}
\Delta(x_{0},\bmx) = \langle\psi\psi^{T}\rangle &\stackrel{\cal P}{\to}& \gamma^{0}\langle\psi\psi^{T}\rangle\gamma^{0} = \gamma^{0}\Delta(x_{0},-\bmx)\gamma^{0}, \\
\bar{\Delta}(x_{0},\bmx) = \langle\bar{\psi}^{T}\bar{\psi}\rangle &\stackrel{\cal P}{\to}& \gamma^{0}\langle\bar{\psi}^{T}\bar{\psi}\rangle\gamma^{0} = \gamma^{0}\bar{\Delta}(x_{0},-\bmx)\gamma^{0}. 
\end{eqnarray}
Under the proper Lorentz transformation ( $\epsilon_{\mu\nu}=-\epsilon_{\nu\mu}$, $\sigma_{\mu\nu}=\frac{i}{2}[\gamma^{\mu},\gamma^{\nu}]$ ),
\begin{eqnarray}
\psi'(x') = S\psi(x), &\quad& \bar{\psi}'(x') = \bar{\psi}(x)S^{-1}, \\
S = \exp(-\frac{i}{4}\epsilon_{\mu\nu}\sigma^{\mu\nu}), &\quad& S^{-1} = \gamma^{0}S^{\dagger}\gamma^{0}, 
\end{eqnarray}
the mean fields are transformed as
\begin{eqnarray}
\Delta'(x') &=& \langle\psi'{\psi^{T}}'\rangle = \langle S\psi\psi^{T}S^{T}\rangle = S\Delta(x)S^{T}  \nonumber \\
&\cong& \bigl(1-\frac{i}{4}\epsilon_{\mu\nu}\sigma^{\mu\nu}\bigr)\Delta(x)\bigl(1-\frac{i}{4}\epsilon_{\mu\nu}(\sigma^{\mu\nu})^{T}\bigr) \nonumber \\
&\cong& \Delta(x)-\frac{i}{4}\epsilon_{\mu\nu}[\sigma^{\mu\nu},\Delta(x)i\gamma^{0}\gamma^{2}]i\gamma^{2}\gamma^{0}, \\
\bar{\Delta}'(x') &\cong& \bar{\Delta}(x)+\frac{i}{4}\epsilon_{\mu\nu}i\gamma^{0}\gamma^{2}[i\gamma^{2}\gamma^{0}\bar{\Delta}(x),\sigma^{\mu\nu}].   
\end{eqnarray}
Here we use $C^{-1}\gamma^{\mu}C=-\gamma^{\mu T}$, $C^{-1}\sigma^{\mu\nu}C=-\sigma^{\mu\nu T}$ and charge conjugation matrix $C=i\gamma^{2}\gamma^{0}$. Then, taking account of the parity, we find we may take scalar as $\gamma^{1}\gamma^{3}=-\gamma_{5}C$ ( $\gamma_{5}=i\gamma^{0}\gamma^{1}\gamma^{2}\gamma^{3}$ ) and pseudo scalar as $\gamma^{0}\gamma^{2}$. We expand the 4$\times$4 matrix mean field into the 16-dimensional complete set of $\gamma$-matrices~[19,34]:
\begin{eqnarray}
\Delta_{\alpha\beta} = \{ \Delta^{S}1 + \Delta^{V}_{\mu}\gamma^{\mu} + \Delta^{T}_{\mu\nu}\sigma^{\mu\nu} + \Delta^{A}_{\mu}\gamma_{5}\gamma^{\mu} + \Delta^{P}i\gamma_{5}\}(-\gamma_{5}C).
\end{eqnarray}
In this expansion, we take a notation that $S$ denotes the scalar, $V$ denotes the vector, $T$ denotes the 2-rank antisymmetric tensor, $A$ denotes the axial vector and $P$ denotes the pseudo scalar. It is clear from our derivation that each set of this expansion is linearly transformed into themselves. This structure of the mean fields in the relativistic theory is one of the essential differences from that of the  nonrelativistic theory, as discussed in Ref. [19]. 

Here we have to mention about the matrix structure of each term given in (36). The scalar, pseudo scalar and vector are antisymmetric with respect to transposition, while other cases are symmetric. These matrix structures have to be related to the Pauli principle, as given in (24). If we want to treat the cases of the axial vector and 2-rank antisymmetric tensor, we have to extend our treatment, and the mean fields have to possess finite angular momentum with odd-parity. This can be treated by introducing the generalized BCS scheme, which will be given in the next section.

Under the charge conjugation, $\psi$ and $\bar{\psi}$ are transformed as   
\begin{eqnarray}
\psi \stackrel{\cal C}{\to} C\bar{\psi}^{T}, \quad \bar{\psi} \stackrel{\cal C}{\to} -\psi^{T}C^{-1},   
\end{eqnarray}  
and hence $\Delta$ and $\bar{\Delta}$ are transformed as
\begin{eqnarray}
\Delta = \langle \psi\psi^{T} \rangle &\stackrel{\cal C}{\to}& C\langle\bar{\psi}^{T}\bar{\psi}\rangle C^{-1} = C\bar{\Delta}C^{-1}, \\
\bar{\Delta} = \langle \bar{\psi}^{T}\bar{\psi} \rangle &\stackrel{\cal C}{\to}& C\langle\psi\psi^{T}\rangle C^{-1} = C\Delta C^{-1}.  
\end{eqnarray}
Under the time reversal, $T=i\gamma^{1}\gamma^{3}$, with the relation
\begin{eqnarray}
\psi(x_{0}) \stackrel{\cal T}{\to} T\psi(-x_{0}), \quad \bar{\psi}(x_{0}) \stackrel{\cal T}{\to} \bar{\psi}(-x_{0})T, 
\end{eqnarray}
together with the rule of taking the complex conjugate about c-numbers, we obtain
\begin{eqnarray}
\Delta = \langle \psi(x_{0})\psi^{T}(x_{0}) \rangle &\stackrel{\cal T}{\to}& -T\langle\psi(-x_{0})\psi^{T}(-x_{0})\rangle^{*}T = -T\Delta^{*}T, \\
\bar{\Delta} = \langle \bar{\psi}^{T}(x_{0})\bar{\psi}(x_{0})\rangle &\stackrel{\cal T}{\to}& -T\langle\bar{\psi}^{T}(-x_{0})\bar{\psi}(-x_{0})\rangle^{*}T = -T(\bar{\Delta})^{*}T. 
\end{eqnarray}
This transformation is related to the concept of ``unitary'' of the mean field~[14]. Throughout this study, we will treat only unitary pairing state. 
 
Chiral transformation can also be treated in the same way, 
\begin{eqnarray}
\psi(x)\to\psi'(x)= e^{i\gamma_{5}\alpha(x)}\psi(x), \quad \bar{\psi}(x)\to\bar{\psi}'(x)= \bar{\psi}(x)e^{i\gamma_{5}\alpha(x)}, 
\end{eqnarray}
then ( $\gamma^{T}_{5}=\gamma_{5}$ ) we obtain
\begin{eqnarray}
\Delta'(x)&=&\langle\psi'(x)\psi^{T'}(x)\rangle = e^{i\gamma_{5}\alpha(x)}\langle\psi(x)\psi^{T}(x)\rangle e^{i\gamma_{5}\alpha(x)} = e^{i\gamma_{5}\alpha(x)}\Delta(x)e^{i\gamma_{5}\alpha(x)}, \\
\bar{\Delta}'(x)&=&\langle\bar{\psi}^{T'}(x)\bar{\psi}'(x)\rangle=e^{i\gamma_{5}\alpha(x)}\langle\bar{\psi}^{T}(x)\bar{\psi}(x)\rangle e^{i\gamma_{5}\alpha(x)}=e^{i\gamma_{5}\alpha(x)}\bar{\Delta}(x)e^{i\gamma_{5}\alpha(x)}. 
\end{eqnarray} 
Thus we will find that, by using the expanded form of the mean field ( Eq. (36) ), the vector and the axial vector are invariant under the chiral transformation. In short, in case of the scalar and pseudo scalar, $\Delta'=e^{2i\gamma_{5}\alpha(x)}\Delta$ holds, in case of the vector and axial vector, $\Delta'_{\mu}=\Delta_{\mu}$ holds, and in case of the antisymmetric tensor, $\Delta'_{\mu\nu}=e^{2i\gamma_{5}\alpha(x)}\Delta_{\mu\nu}$ holds.

\section{The Generalized BCS Theory}  
\label{sec:gen}

In this section, to treat especially the spin triplet Cooper pair states, we investigate the method of the generalized BCS approach in the relativistic theory. Like the nonrelativistic theory, our starting point of studying the generalized BCS theory is a Gor'kov equation with generalized pairing scheme. We introduce next generalized Gor'kov equations:  
\begin{eqnarray}
(i\gamma^{\mu}\partial_{\mu}-m+\gamma^{0}\mu)_{x}S_{F}(x,y)-i\int d^{4}z V(x,z)F(x,z)\bar{F}(z,y) = \delta^{(4)}(x-y), \\
(i\gamma^{\mu}\partial_{\mu}-m+\gamma^{0}\mu)_{x}F(x,y)-i\int d^{4}z V(x,z)F(x,z)S_{F}(y,z) = 0, \\
(i\gamma^{\mu T}\partial_{\mu}+m-\gamma^{0T}\mu)_{x}\bar{F}(x,y)+i\int d^{4}z V(x,z)F(x,z)S_{F}(z,y)^{T} = 0, \\
(i\gamma^{\mu T}\partial_{\mu}+m-\gamma^{0T}\mu)_{x}(-)S_{F}(y,x)^{T}-i\int d^{4}z V(x,z)\bar{F}(x,z)F(z,y) = \delta^{(4)}(x-y).  
\end{eqnarray}
In the Fourier transform in a matrix form, they are expressed as
\begin{eqnarray}
\left(
\begin{array}{cccc}
\kfey -m+\gamma^{0}\mu & \Delta(k) \\
\bar{\Delta}(k) & \kfey^{T} +m-\gamma^{0T}\mu 
\end{array}
\right)
\left( 
\begin{array}{cccc}
S_{F}(k) & -iF(k) \\
-i\bar{F}(k) & -S_{F}(-k)^{T} 
\end{array}
\right)
=
\left(
\begin{array}{cccc}
1 & 0 \\
0 & 1 
\end{array}
\right),
\end{eqnarray}
where the definition of the mean fields are
\begin{eqnarray}
\Delta(k) &=& \int\frac{d^{4}p}{(2\pi)^{4}}V(k,p)F(p), \\
\bar{\Delta}(k) &=& \int\frac{d^{4}p}{(2\pi)^{4}}V(k,p)\bar{F}(p).
\end{eqnarray}
Here $V(k,p)$ is an effective attractive interaction which is assumed to give rise to superconducting states. Here we consider $V(k,p)$ as a scalar function. We can assume it as a Feynman propagator for a scalar field. The definite form of the function $V(k,p)$ is related to the symmetry of the Cooper pair; different symmetries correspond to the different forms of $V(k,p)$. Since the matrix structure of (50) is same as (16), the Green's function obtained by solving (50) gives the same matrix structure. It is also clear that this can be extended to the finite-temperature Matsubara formalism. 

Next, we decompose the interaction and the mean fields into each channel. When we denote the total momentum of a Cooper pair as $P_{\mu}=(P_{0},\bmP), P^{2}=P^{2}_{0}-\bmP^{2}$, and when $P^{2}>0$, we can always stand on a rest frame which satisfies the condition $\bmP=0$. We treat our problem under the condition $P_{0}>0, \bmP=0$. Our mean fields have the translation invariance. At that time the Wigner little group of the Poincar\'{e} group becomes $O(3)$ rotation~[34]. It should be enough to use the irreducible representation of $O(3)$, i.e. the three dimensional spherical harmonics, for the channel decomposition. It is also related to the fact that our theory will mainly treat the finite-temperature Matsubara formalism, which selects a specific time-coordinate. In this formalism, the theory allows only $O(3)$ rotational symmetry. Under such consideration, we assume that the interaction depends only on the angle between $\bmk$ and $\bmp$, expanding it by using the addition theorem,      
\begin{eqnarray}
V(k,p) &=& \sum_{l}\sum_{m_{l}}(2l+1)V_{l}(k_{0},|\bmk|;p_{0},|\bmp|)Y_{lm_{l}}(\hat{\bmk})Y^{*}_{lm_{l}}(\hat{\bmp}) \nonumber \\
&\to& \sum_{l}\sum_{m_{l}}4\pi g_{l}Y_{lm_{l}}(\hat{\bmk})Y^{*}_{lm_{l}}(\hat{\bmp}),  
\end{eqnarray}
where we also introduce the weak coupling approximation $V_{l}(k_{0},|\bmk|;p_{0},|\bmp|)\approx g_{l}$~[35]. This approximation neglects the retardation of the interaction and also neglects the dependence on the magnitude of each momentum. We may call this as ``anisotropic contact'' interaction. If one of these channels is attractive, the Fermi sea can be unstable and it can become a superconducting state.

Later we will find that the gap equations which can have nontrivial solutions are those for the scalar $\Delta^{S}$, the 0th component of vector $\Delta^{V}_{0}$, the space-like components of axial vector ${\bf \Delta}^{A}$, the axial-vector-like components of 2-rank antisymmetric tensor ${\bf \Delta}^{T}_{(A)}$. We have to discuss here only these mean fields. To satisfy the Pauli principle, for the case of spin singlet pairing, we select the parity-even part of the expansion of the interaction (53),  
\begin{eqnarray}
V^{(e)}(k,p) &=& g_{0}+g_{2}\sum_{m=2}^{-2}Y_{2m}(\hat{\bmk})Y^{*}_{2m}(\hat{\bmp}) + \cdots.  
\end{eqnarray}
Here the first term corresponds to the interaction $g_{0}(\bar{\psi}\psi)^{2}/2$. The mean field is also expanded in the same way,
\begin{eqnarray}
\Delta^{1} &=& \Delta_{0}+\sum_{m=2}^{-2}\Delta_{2m}Y_{2m}(\bmk)+\cdots.
\end{eqnarray}
On the other hand, for the spin triplet pairing, we have to select the parity-odd part of the expansion. As for the interaction, we have 
\begin{eqnarray}
V^{(o)}(k,p) &=& g_{1}\sum_{m=1}^{-1}Y_{1m}(\hat{\bmk})Y^{*}_{1m}(\hat{\bmp})+g_{3}\sum_{m=3}^{-3}Y_{3m}(\hat{\bmk})Y^{*}_{3m}(\hat{\bmp})+\cdots. 
\end{eqnarray}
We need some consideration for the treatment of the spin triplet mean fields. Because of the relativistic nature of the theory, spin and orbital degrees of freedom can not be treated separately. The spin projection to the space-fixed z-axis is not conserved. Now the good quantum numbers are the total angular momentum $j$, its projection $m_{j}$ and helicity $\lambda$. We will treat the spin triplet Cooper pair state as an axial vector boson, with even internal parity. The basis for the expansion of the triplet mean field then becomes the helicity states. We define the space-fixed coodinate as the $k_{x}k_{y}k_{z}$-system, while the body-fixed coordinate moving with the boson as the $\xi\eta\zeta$-system. We choose the direction $\hat{\bmk}$ along the $\zeta$-axis. The helicity is the projection to this axis, $\lambda=-1,+1$(transverse), $\lambda=0$(longitudinal). Here we take the phase convention after the textbook of Landau and Lifshitz~[36]. Then the basis becomes
\begin{eqnarray}
\psi_{jm\lambda}(\hat{\bmk}) &=& i^{j-1}\sqrt{\frac{2j+1}{4\pi}}{\bf e}^{(\lambda)}{\cal D}^{(j)}_{\lambda m}(\hat{\bmk}), 
\end{eqnarray}     
where ${\bf e}^{(\lambda)}$ is the spherical unit vector of the $\xi\eta\zeta$-system,
\begin{eqnarray}
{\bf e}^{(\pm1)} = \mp\frac{i}{\sqrt{2}}(\hat{\xi}\pm i\hat{\eta}), \quad {\bf e}^{(0)} = i\hat{\zeta}, 
\end{eqnarray}
As for the Wigner rotation matrix, we determine the relation between the Eulerian angles and spherical angles of the $k_{x}k_{y}k_{z}$-system as
\begin{eqnarray}
{\cal D}^{(j)}_{\lambda m}(\alpha,\beta,\gamma) &=& {\cal D}^{(j)}_{\lambda m}(\phi,\theta,0) = e^{im\phi}d^{(j)}_{\lambda m}(\theta). 
\end{eqnarray}
The orthonormal relation is given from the equations,
\begin{eqnarray}
{\bf e}^{(\lambda_{1})*}\cdot{\bf e}^{(\lambda_{2})} &=& \delta_{\lambda_{1}\lambda_{2}}, \\
\int^{\pi}_{0}\sin\theta d\theta\int^{2\pi}_{0}d\phi{\cal D}^{(j_{1})*}_{\lambda_{1}m_{1}}(\hat{\bmk}){\cal D}^{(j_{2})}_{\lambda_{2}m_{2}}(\hat{\bmk}) &=& \frac{4\pi}{2j_{1}+1}\delta_{j_{1}j_{2}}\delta_{m_{1}m_{2}}, 
\end{eqnarray}
as
\begin{eqnarray}
\int^{\pi}_{0}\sin\theta d\theta\int^{2\pi}_{0}d\phi \psi^{*}_{j_{1}m_{1}\lambda_{1}}(\hat{\bmk})\cdot\psi_{j_{2}m_{2}\lambda_{2}}(\hat{\bmk}) &=& \delta_{j_{1}j_{2}}\delta_{m_{1}m_{2}}\delta_{\lambda_{1}\lambda_{2}}.   
\end{eqnarray}
About the inversion $\alpha\equiv\phi\to\phi+\pi$, $\beta\equiv\theta\to\pi-\theta$, $\gamma\to\pi-\gamma$, each function is transformed as
\begin{eqnarray}
{\cal D}^{(j)}_{\lambda m}(\hat{\bmk}) &=& (-1)^{j-\lambda}{\cal D}^{(j)}_{-\lambda m}(-\hat{\bmk}), \\
{\bf e}^{(\lambda)}(\hat{\bmk}) &=& (-1)^{1-\lambda}{\bf e}^{(-\lambda)}(-\hat{\bmk}). 
\end{eqnarray}  
Thus the parity eigenstates become
\begin{eqnarray}
\psi_{jm0}(\hat{\bmk}) &=& i^{j-1}\sqrt{\frac{2j+1}{4\pi}}{\bf e}^{(0)}{\cal D}^{(j)}_{0m}(\hat{\bmk}),  \\
\psi^{(+)}_{jm|\lambda|}(\hat{\bmk}) &=& i^{j-1}\sqrt{\frac{2j+1}{8\pi}}({\bf e}^{(1)}{\cal D}^{(j)}_{1m}(\hat{\bmk})+{\bf e}^{(-1)}{\cal D}^{(j)}_{-1m}(\hat{\bmk})), \\
\psi^{(-)}_{jm|\lambda|}(\hat{\bmk}) &=& i^{j-1}\sqrt{\frac{2j+1}{8\pi}}({\bf e}^{(1)}{\cal D}^{(j)}_{1m}(\hat{\bmk})-{\bf e}^{(-1)}{\cal D}^{(j)}_{-1m}(\hat{\bmk})).
\end{eqnarray}
Each parity is given as; $\psi_{jm0}:(-1)^{j+1}$, $\psi^{(+)}_{jm|\lambda|}:(-1)^{j+1}$, $\psi^{(-)}_{jm|\lambda|}:(-1)^{j}$. Thus the odd-parity states becomes $\psi_{000}$ for the monopole ($j=0$), $\psi^{(-)}_{1m1}$ for the dipole ($j=1$), $\psi_{2m0}$ and $\psi^{(+)}_{2m1}$ for the quadrepole, etc. Since in this work, we treat only the unitary states of the mean fields, and $({\bf \Delta}\cdot\sigma)({\bf \Delta}^{*}\cdot\sigma)={\bf \Delta}\cdot{\bf \Delta}^{*}+i({\bf \Delta}\times{\bf \Delta}^{*})\cdot\sigma$, we have the unitary condition ${\bf \Delta}\times{\bf \Delta}^{*}=0$~[14]. The ``longitudinal'' $\psi_{jm0}$ is itself unitary, but as for the ``transverse'' $\psi^{(\pm)}_{jm1}$, we have to add the complex conjugate to itself. This means we have to take a linear combination with the time-reversal state in order to obtain a unitary state. Using the relations
\begin{eqnarray}
\psi^{(+)*}_{jm1} &=& (-1)^{j-m}\psi^{(+)}_{j-m1}, \nonumber \\
\psi^{(-)*}_{jm1} &=& (-1)^{j-m+1}\psi^{(-)}_{j-m1}, 
\end{eqnarray}
we obtain the following bases that we use for the expansion:
\begin{eqnarray}
& & \psi_{jm0}, \\
\frac{1}{\sqrt{2}}[\psi^{(+)}_{jm1}&+&(-1)^{j-m}\psi^{(+)}_{j-m1}], \\
\frac{1}{\sqrt{2}}[\psi^{(-)}_{jm1}&-&(-1)^{j-m}\psi^{(-)}_{j-m1}].
\end{eqnarray}
In the ``longitudinal'' basis, the direction of the mean field vector coincides with $\hat{\bmk}$, while in the ``transversal'' basis, it is orthogonal to $\hat{\bmk}$. Finally the spin triplet mean fields are expanded in the following form:
\begin{eqnarray}
{\bf \Delta}(k) &=& \sum_{jm_{j}} \Delta^{(0)}_{jm_{j}} \psi_{jm_{j}0} (\hat{\bmk}),  \\       
{\bf \Delta}(k) &=& \sum_{jm_{j}} \Delta^{(+)}_{jm_{j}} \frac{1}{\sqrt{2}} (\psi^{(+)}_{jm_{j}1} (\hat{\bmk}) + (-1)^{j-m_{j}} \psi^{(+)}_{j-m_{j}1} (\hat{\bmk}) ),  \\
{\bf \Delta}(k) &=& \sum_{jm_{j}} \Delta^{(-)}_{jm_{j}} \frac{1}{\sqrt{2}} (\psi^{(-)}_{jm_{j}1} (\hat{\bmk}) - (-1)^{j-m_{j}} \psi^{(-)}_{j-m_{j}1} (\hat{\bmk})),
\end{eqnarray}
where we have to select only the parity-odd part of the expansion, for satisfing the Pauli principle. Here we also take the so-called weak coupling approximation $\Delta_{jm_{j}}(k_{0},|\bmk|)\approx\Delta_{jm_{j}}$~[35].

\section{The Solutions for the Homogeneous System}
\label{sec:sol}

Now using the definition $C^{-1}\gamma^{\mu}C=-\gamma^{\mu T}$ and the notation $\tilde{k}\equiv(k_{0}+\mu,\bmk)$, $\check{k}\equiv(k_{0}-\mu,\bmk)$, our Gor'kov equation becomes
\begin{eqnarray}
(\tilde{\kfey}-m)S_{F}(k)-i\Delta\bar{F}(k) = 1, \\
(\tilde{\kfey}-m)F(k)-i\Delta S_{F}(-k)^{T} = 0, \\
-C^{-1}(\check{\kfey}-m)C\bar{F}(k)+i\bar{\Delta}S_{F}(k) = 0, \\
-C^{-1}(\check{\kfey}-m)C(-)S_{F}(-k)^{T}-i\bar{\Delta}F(k) = 1.
\end{eqnarray}
First, $S_{F}(k)$ and $\bar{F}(k)$ satisfy the relation
\begin{eqnarray}
\bar{F}(k) &=& iC^{-1}\frac{\check{\kfey}+m}{\check{k}^{2}-m^{2}}C\bar{\Delta}S_{F}(k).
\end{eqnarray}
Then we obtain
\begin{eqnarray}
(\tilde{\kfey}-m)S_{F}(k) + \Delta C^{-1}\frac{\check{\kfey}+m}{\check{k}^{2}-m^{2}}C\bar{\Delta}S_{F}(k) = 1. 
\end{eqnarray}
On the other hand, $S_{F}(-k)^{T}$ and $F(k)$ satisfy 
\begin{eqnarray}
F(k) &=& i\frac{\tilde{\kfey}+m}{\tilde{k}^{2}-m^{2}}\Delta S_{F}(-k)^{T}, 
\end{eqnarray}
then
\begin{eqnarray}
(\check{\kfey}-m)(CS_{F}(-k)^{T}C^{-1})+C\bar{\Delta}\frac{\tilde{\kfey}+m}{\tilde{k}^{2}-m^{2}}\Delta C^{-1}(CS_{F}(-k)^{T}C^{-1}) &=& 1.
\end{eqnarray}
Therefore, our problem turns out to solve the two 4$\times$4 matrix equations (80) and (82). Unfortunately, it is difficult to solve these equations completely because of the matrix structure of the Dirac operator, particularly in the analytical form. We have to solve the equations assuming the type of the mean field that might be realized. We use the expanded form of the mean field,
\begin{eqnarray}
\Delta &=& \{\Delta^{S}+\Delta^{P}i\gamma_{5}+\Delta^{V}_{0}\gamma^{0}+\Delta^{A}_{0}\gamma_{5}\gamma^{0} \nonumber \\
& & +{\bf \Delta}^{V}\cdot\vec{\gamma}+{\bf \Delta}^{A}\cdot\gamma_{5}\vec{\gamma}+{\bf \Delta}^{T}_{(V)}\cdot i\gamma^{0}\vec{\gamma}+{\bf \Delta}^{T}_{(A)}\cdot\gamma_{5}\gamma^{0}\vec{\gamma}\}(-\gamma_{5}C),
\end{eqnarray}
and put each type of the mean field separately into the Gor'kov equations.
In this way, we can treat each cases of the mean fields in the same manner. In particular, the case of the vector mean fields, under the unitary assumption, can be treated completely in the same way as the scalar mean field case. We write down the expanded form of the mean field ( Eq. (83) ) as, 
\begin{eqnarray}
\Delta C^{-1} &=& -\{\sum^{8}_{a=1}\Delta_{a}\cdot\Gamma^{a}\}\gamma_{5}, \\
C\bar{\Delta} &=& -\gamma_{5}\{\sum^{8}_{a=1}\Delta^{*}_{a}\cdot\Gamma^{a}\}. 
\end{eqnarray}
Going back to (80) and (82), puting one of 8 types of mean fields $\Delta_{a}$, under the unitary assumption, $(\Delta_{a}\cdot\Gamma^{a})(\Delta^{*}_{a}\cdot\Gamma^{a})=(\Delta^{*}_{a}\cdot\Gamma^{a})(\Delta_{a}\cdot\Gamma^{a})=\pm(\Delta_{a}\cdot\Delta^{*}_{a})$ ( where plus sign corresponds to the scalar, 0th component of vector, space-like components of axial vector and axial-vector-like components of 2-rank antisymmetric tensor, while minus sign corresponds to the other cases ), and also using $\gamma_{5}\Gamma^{a} = \pm\Gamma^{a}\gamma_{5}$ ( plus sign corresponds to the scalar, pseudo scalar and antisymmetric tensor, while minus sign corresponds to the vector and axial vector ), we obtain
\begin{eqnarray}
(\tilde{\kfey}-m)S_{F}(k)-(\Delta_{a}\cdot\Gamma^{a})\frac{\check{\kfey}-m}{\check{k}^{2}-m^{2}}(\Delta^{*}_{a}\cdot\Gamma^{a})S_{F}(k) &=& 1, \\
(\check{\kfey}-m)CS_{F}(-k)^{T}C^{-1}-(\Delta^{*}_{a}\cdot\Gamma^{a})\frac{\tilde{\kfey}-m}{\tilde{k}^{2}-m^{2}}(\Delta_{a}\cdot\Gamma^{a})CS_{F}(-k)^{T}C^{-1} &=& 1.
\end{eqnarray}
Solving this equations, we obtain
\begin{eqnarray}
S_{F}(k) &=& \frac{1}{D(k)}\{(\check{k}^{2}-m^{2})(\tilde{\kfey}+m)-(\Delta_{a}\cdot\Gamma^{a})(\check{\kfey}+m)(\Delta^{*}_{a}\cdot\Gamma^{a})\}, \nonumber \\
-iF(k) &=& (\pm)\frac{1}{D(k)}\gamma_{5}\{(\tilde{\kfey}-m)(\Delta_{a}\cdot\Gamma^{a})(\check{\kfey}+m)-(\pm)(\Delta_{a}\cdot\Delta^{*}_{a})(\Delta_{a}\cdot\Gamma^{a})\}C,  \nonumber \\    
-i\bar{F}(k) &=& \frac{1}{D(k)}C^{-1}\gamma_{5}\{(\check{\kfey}-m)(\Delta^{*}_{a}\cdot\Gamma^{a})(\tilde{\kfey}+m)-(\pm)(\Delta_{a}\cdot\Delta^{*}_{a})(\Delta^{*}_{a}\cdot\Gamma^{a})\}, \nonumber \\    
-S_{F}(-k)^{T} &=& -\frac{1}{D(k)}C^{-1}\{(\tilde{k}^{2}-m^{2})(\check{\kfey}+m)-(\Delta^{*}_{a}\cdot\Gamma^{a})(\tilde{\kfey}+m)(\Delta_{a}\cdot\Gamma^{a})\}C, \nonumber \\  
D(k) &=& (\tilde{k}^{2}-m^{2})(\check{k}^{2}-m^{2}) \nonumber \\
 & & {} -\{\tilde{\kfey}(\Delta_{a}\cdot\Gamma^{a})\check{\kfey}(\Delta^{*}_{a}\cdot\Gamma^{a})+(\Delta_{a}\cdot\Gamma^{a})\check{\kfey}(\Delta^{*}_{a}\cdot\Gamma^{a})\tilde{\kfey}-2m^{2}(\pm)(\Delta_{a}\cdot\Delta^{*}_{a})\} \nonumber \\
& & {} +(\Delta_{a}\cdot\Delta^{*}_{a})^{2}. 
\end{eqnarray}
As for the sign for $F(k)$, plus corresponds to the scalar, pseudo scalar and antisymmetric tensor, while minus corresponds to the vector and axial vector. All the eight Green's functions we have obtained reflect the features of each type of the mean fields, both in the numerator and denominator. We can confirm that this solution satisfies the Gor'kov equation. We can see that when we take the limit $\Delta_{a}\to 0$, the solution gives the Green's functions of normal state. 

We have solved the Gor'kov equations under the assumption of specific types of the mean fields in order to make them tractable. We have to check the existence of coupling among different types of the mean fields. For this purpose, we take the gap equations
\begin{eqnarray}
\Delta &=& g_{0}\int\frac{d^{4}p}{(2\pi)^{4}}F(p), \\
\Delta(k) &=& \int\frac{d^{4}p}{(2\pi)^{4}}V(k,p)F(p).
\end{eqnarray}
We substitute the expanded form of $\Delta$ ( Eq. (83) ) into the left hand side, substitute the solutions for $F(p)$ obtained previously into the right hand side, and take the trace of both side. We find couplings between $\Delta^{S}$ and $\Delta^{V}_{0}$, $\Delta^{P}$ and $\Delta^{A}_{0}$, ${\bf \Delta}^{V}$ and ${\bf \Delta}^{T}_{(V)}$, ${\bf \Delta}^{A}$ and ${\bf \Delta}^{T}_{(A)}$. We should say that the Green's functions obtained above are approximate ones by neglecting these couplings, and also the gap equations obtained by using these Green's functions contain these approximations. 

In the end of this section, we investigate the factorization of the denominators of the Green's functions. We discuss some features of quasiparticle dispersion. In all of the cases, the denominator given in (88) becomes the following form: 
\begin{eqnarray}
D(k) &=& (\tilde{k}^{2}-m^{2})(\check{k}^{2}-m^{2})-2(\Delta_{a}\cdot\Delta^{*}_{a})(\tilde{k}\cdot\check{k}+d)+(\Delta_{a}\cdot\Delta^{*}_{a})^{2}. 
\end{eqnarray}
This is 2nd order in $k^{2}_{0}$, thus easily factorize it as
\begin{eqnarray}
D(k) &=& (k_{0}-E_{+})(k_{0}+E_{+})(k_{0}-E_{-})(k_{0}+E_{-}).
\end{eqnarray}
$D(k)$ is fourth order in $k_{0}$, but the Gor'kov equation itself is 8$\times$8 matrix. Like the nonrelativistic $^{1}S$-BCS theory, the quasiparticle spectra are doubly degenerate. We will find that the condition for doubly degenerated quasiparticle spectra is the unitary of the mean fields, which implies the time reversal symmetry of the mean fields~[14]. $E_{+}$ is the branch for the quasiparticle coming from the positive energy solution, while $E_{-}$ is the branch for quasiparticle coming from the negative energy solution. The expression for the pole becomes 
\begin{eqnarray}
E^{2}_{\pm} &=& \bmk^{2}+m^{2}+\mu^{2}+(\Delta_{a}\cdot\Delta^{*}_{a})\mp\sqrt{4\mu^{2}(\bmk^{2}+m^{2})+2(\Delta_{a}\cdot\Delta^{*}_{a})(m^{2}+d),}
\end{eqnarray}
where $d=-m^{2}$ for the scalar, $d=+m^{2}$ for the pseudo scalar, $d=2\bmk^{2}-m^{2}$ for the 0th component of vector, $d=2|\bmk\cdot{\bf \Delta}^{V}|^{2}/({\bf \Delta}^{V}\cdot{\bf \Delta}^{V*})+m^{2}$ for the space-like components of vector, $d=2\bmk^{2}+m^{2}$ for the 0th component of axial vector, $d=2|\bmk\cdot{\bf \Delta}^{A}|^{2}/({\bf \Delta}^{A}\cdot{\bf \Delta}^{A*})-m^{2}$ for the space-like components of axial vector, $d=-2|\bmk\cdot{\bf \Delta}^{T}_{(V)}|^{2}/({\bf \Delta}^{T}_{(V)}\cdot{\bf \Delta}^{T*}_{(V)})+2\bmk^{2}+m^{2}$ for the vector-like components of 2-rank antisymmetric tensor and $d=-2|\bmk\cdot{\bf \Delta}^{T}_{(A)}|^{2}/({\bf \Delta}^{T}_{(A)}\cdot{\bf \Delta}^{T}_{(A)})+2\bmk^{2}-m^{2}$ for the axial-vector-like components of the 2-rank antisymmetric tensor.  About the $d$, the sign before $m^{2}$ is coming from the parity of the mean fields which is determined by $\gamma$-matrices, minus for even parity and plus for odd parity. It holds that in the limit $m\to 0$, the quasiparticle spectra for the scalar and pseudo scalar, 0th component of the vector and axial vector, space-like components of the vector and axial vector, vector-like components and axial-vector-like components of the 2-rank antisymmetric tensor, coinside. We write the quasiparticle dispersion explicitly. For the scalar case, as obtaied in Ref. ~[18],  
\begin{eqnarray}
E^{2}_{\pm} &=& (\sqrt{\bmk^{2}+m^{2}}\mp\mu)^{2}+|\Delta^{S}|^{2}. 
\end{eqnarray} 
In the pseudo scalar case,
\begin{eqnarray}
E^{2}_{\pm} &=& \bmk^{2}+m^{2}+\mu^{2}+|\Delta_{a}|^{2}\mp2\sqrt{(\bmk^{2}+m^{2})\mu^{2}+m^{2}|\Delta^{P}|^{2}},
\end{eqnarray}
and these spectra coinside at $m\to 0$, $E^{2}_{\pm} = (|\bmk|\mp\mu)^{2}+|\Delta|^{2}$. In the case of the 0th component of the vector pairing, 
\begin{eqnarray}
E^{2}_{\pm} &=& \bmk^{2}+m^{2}+\mu^{2}+|\Delta^{V}_{0}|^{2}\mp 2 \sqrt{(\bmk^{2}+m^{2})\mu^{2}+|\Delta^{V}_{0}|^{2}\bmk^{2}},  
\end{eqnarray}
and in the limit $m\to 0$, $E^{2}_{\pm} = (|\bmk|\mp\sqrt{\mu^{2}+|\Delta^{V}_{0}|^{2}})^{2}$. In the case of the space-like components of the vector pairing, we obtain
\begin{eqnarray}
E^{2}_{\pm} &=& \bmk^{2}+m^{2}+\mu^{2}+({\bf \Delta}^{V}\cdot{\bf \Delta}^{V*})\mp2\sqrt{(\bmk^{2}+m^{2})\mu^{2}+m^{2}({\bf \Delta}^{V}\cdot{\bf \Delta}^{V*})+|\bmk\cdot{\bf \Delta}^{V}|^{2}}.    \nonumber \\
& &  
\end{eqnarray}
In the case of the 0th component of the axial vector pairing,
\begin{eqnarray}
E^{2}_{\pm} &=& (\sqrt{\bmk^{2}+m^{2}}\mp\sqrt{\mu^{2}+|\Delta^{A}_{0}|^{2}})^{2},
\end{eqnarray}
and in the limit $m\to 0$, the cases of the 0th component of the vector and axial vector coincide with each other. The case for the space-like components of the axial vector pairing, we obtain
\begin{eqnarray}
E^{2}_{\pm} &=& \bmk^{2}+m^{2}+\mu^{2}+({\bf \Delta}^{A}\cdot{\bf \Delta}^{A*})\mp2\sqrt{(\bmk^{2}+m^{2})\mu^{2}+|\bmk\cdot{\bf \Delta}^{A}|^{2}}, \nonumber \\
& & 
\end{eqnarray}
and for $m\to 0$, this also coinsides with the case of the space-like components of the vector pairing. About the vector-like components of the 2-rank antisymmetric tensor, we obtain 
\begin{eqnarray}
E^{2}_{\pm} &=& \bmk^{2}+m^{2}+\mu^{2}+({\bf \Delta}^{T}_{(V)}\cdot{\bf \Delta}^{T*}_{(V)})\mp\sqrt{(\bmk^{2}+m^{2})(\mu^{2}+{\bf \Delta}^{T}_{(V)}\cdot{\bf \Delta}^{T*}_{(V)})-|\bmk\cdot{\bf \Delta}^{T}_{(V)}|^{2}}, \nonumber \\
& &          
\end{eqnarray}
and the axial-vector-like components of the 2-rank antisymmetric tensor becomes
\begin{eqnarray}
E^{2}_{\pm} &=& \bmk^{2}+m^{2}+\mu^{2}+({\bf \Delta}^{T}_{(A)}\cdot{\bf \Delta}^{T*}_{(A)})\mp2\sqrt{(\bmk^{2}+m^{2})\mu^{2}+\bmk^{2}({\bf \Delta}^{T}_{(A)}\cdot{\bf \Delta}^{T*}_{(A)})-|\bmk\cdot{\bf \Delta}^{T}_{(A)}|^{2}}. \nonumber \\
& & 
\end{eqnarray}
These spectra also coinside with each other in the limit $m\to 0$.

The dispersion of the quasiparticle includes a gap topology, and it reflects orbital part of the mean field. It is clear that the complexity of the dispersion in relativistic case is the result of the matrix structure of the Dirac operator. We can make an assertion that these situations manifest the fact that in the relativistic theory, we cannot decouple the spin and orbital degrees of freedom. The dispersion and also the gap topology reflect the density of states and then affect the thermodynamics like the specific heat. They influence also the collective modes. Spin part of the mean field gives spin symmetry of the Cooper pair, and it is reflected to the spin susceptibility or the Knight shift. In all the cases given above, we obtain $E^{2}_{\pm}=(\sqrt{\bmk^{2}+m^{2}}\mp\mu)^{2}$ in the limit $\Delta_{a}\to 0$.

\section{The Gap Equations}
\label{sec:gap}

In this section, we will obtain gap equations by using our Green's functions obtained previously. We use the finite-temperature Matsubara formalism. The gap function determines the thermodynamic properties of the superconductivity as direct consequences. The gap equation itself is the stationary condition of the free energy~[10]. 

Under the Pauli principle, and with the condition $g_{l}>0$, we will find that we have nontrivial solutions for the mean fields in the case of the scalar and 0th component of vector for even-parity interaction $V^{(e)}$, the space-like components of axial vector, and axial-vector-like components of 2-rank antisymmetric tensor for odd-parity interaction $V^{(o)}$. Other cases give the gap equation of the form ``$1=$ negative value'', thus under $g_{l}>0$, there is no nontrivial solution. Thus in the interaction $g_{0}(\bar{\psi}\psi)^{2}/2$ with $g_{0}>0$, it can give nontrivial solutions only in the case of the scalar pairing and 0th component of vector pairing.

First, we deel with the interaction $g_{0}(\bar{\psi}\psi)^{2}/2$. The self-consistency condition becomes
\begin{eqnarray}
\Delta &=& g_{0}\sum_{n}\frac{1}{\beta}\int\frac{d^{3}\bmk}{(2\pi)^{3}}{\cal F}(\omega_{n},\bmk),  \\
\bar{\Delta} &=& g_{0}\sum_{n}\frac{1}{\beta}\int\frac{d^{3}\bmk}{(2\pi)^{3}}\bar{\cal F}(\omega_{n},\bmk).  
\end{eqnarray}
Substitute ${\cal F}$ for scalar or 0th component of vector into (102), deal with the $\gamma$-matrices and take trace of both sides. Hereafter we only deal with the nontrivial case $\Delta_{a}\ne 0$. In the denominator $D(\omega_{n},\bmk)=(i\omega_{n}-E_{+})(i\omega_{n}+E_{+})(i\omega_{n}-E_{-})(i\omega_{n}+E_{-})$ ( substitute $k_{0}\to i\omega_{n}$ in (92) for the Matsubara formalism ), we perform the partial fraction decomposition and discrete frequency summation. For the scalar case, we obtain the gap equation of the same form as derived by Strange~[26]: 
\begin{eqnarray}
1 &=& \frac{g_{0}}{2}\int^{\Lambda}_{-\Lambda}\frac{d^{3}\bmk}{(2\pi)^{3}}\Bigl(\frac{1}{2E_{+}}\tanh\frac{\beta}{2}E_{+} + \frac{1}{2E_{-}}\tanh\frac{\beta}{2}E_{-}\Bigr), \\
E_{\pm} &=& \sqrt{(\sqrt{\bmk^{2}+m^{2}}\mp\mu)^{2}+|\Delta^{S}|^{2}}.
\end{eqnarray}
Here we introduce the cut-off for the momentum integration. Of course, this cut-off scheme can not be a covariant form. In the high-energy physics, there are problems that a different cut-off scheme gives different physical quantities~[4,5]. But we do not treat these problems here. Eqs. (104) and (105) give simple extension of the nonrelativistic $^{1}S$-BCS theory. The second term of the integrand in (104) is the contribution of the quasiparticle coming from the negative energy solution. 

In the case of the 0th component of the vector pairing, the gap equation is given in the following form:  
\begin{eqnarray}
1 &=& \frac{g_{0}}{2}\int\frac{d^{3}\bmk}{(2\pi)^{3}}\Bigl(\{1-\frac{\bmk^{2}}{\sqrt{(\bmk^{2}+m^{2})\mu^{2}+|\Delta^{V}_{0}|^{2}\bmk^{2}}}\}\frac{1}{2E_{+}}\tanh\frac{\beta}{2}E_{+} \nonumber \\
& & +\{1+\frac{\bmk^{2}}{\sqrt{(\bmk^{2}+m^{2})\mu^{2}+|\Delta^{V}_{0}|^{2}\bmk^{2}}}\}\frac{1}{2E_{-}}\tanh\frac{\beta}{2}E_{-}\Bigr),  \\
E_{\pm} &=& \sqrt{\bmk^{2}+m^{2}+\mu^{2}+|\Delta^{V}_{0}|^{2}\mp2\sqrt{(\bmk^{2}+m^{2})\mu^{2}+|\Delta^{V}_{0}|^{2}\bmk^{2}}}. 
\end{eqnarray}

On the other hand, in the case of the pseudo scalar pairing and 0th component of the axial vector pairing, the gap equations become
\begin{eqnarray}
-1 &=& \frac{g_{0}}{2}\int\frac{d^{3}\bmk}{(2\pi)^{3}}\Bigl(\Bigl\{1-\frac{m^{2}}{\sqrt{(\bmk^{2}+m^{2})\mu^{2}+|\Delta^{P}|^{2}m^{2}}}\Bigr\}\frac{1}{2E_{+}}\tanh\frac{\beta}{2}E_{+} \nonumber \\
& & +\Bigl\{1+\frac{m^{2}}{\sqrt{(\bmk^{2}+m^{2})\mu^{2}+|\Delta^{P}|^{2}m^{2}}}\Bigr\}\frac{1}{2E_{-}}\tanh\frac{\beta}{2}E_{-}\Bigr), \\
E_{\pm} &=& \sqrt{\bmk^{2}+m^{2}+\mu^{2}+|\Delta^{P}|^{2}\mp 2\sqrt{(\bmk^{2}+m^{2})\mu^{2}+|\Delta^{P}|^{2}m^{2}}},
\end{eqnarray}
and
\begin{eqnarray}
-1 &=& \frac{g_{0}}{2}\int\frac{d^{3}\bmk}{(2\pi)^{3}}\Bigl(-\frac{1}{2\sqrt{\mu^{2}+|\Delta^{A}_{0}|^{2}}}\tanh\frac{\beta}{2}E_{+}+\frac{1}{2\sqrt{\mu^{2}+|\Delta^{A}_{0}|^{2}}}\tanh\frac{\beta}{2}E_{-}\Bigr), \\
E_{\pm} &=& \sqrt{\bmk^{2}+m^{2}}\mp\sqrt{\mu^{2}+|\Delta^{A}_{0}|^{2}}. 
\end{eqnarray}
We find that the right hand side of each equations are positive. Hence, we do not obtain nontrivial solutions in these cases. Due to the same reason, the cases for the space-like components of the vector pairing and the vector-like components of the 2-rank antisymmetric tensor pairing also have no nontrivial solutions for $g_{l}>0$.

Next, we treat the spin triplet states by using the generalized BCS formalism. Adopting  the Green's functions of the space-like components of the axial vector or the axial-vector-like components of the 2-rank antisymmetric tensor for the self-consistency condition, and manipulating the $\gamma$-matrices, we obtain a 3-vector gap equation: 
\begin{eqnarray}
{\bf \Delta}(\bmk) &=& \sum_{n}\frac{1}{\beta}\int \frac{d^{3}\bmp}{(2\pi)^{3}}V^{(o)}(\bmk, \bmp)\frac{1}{D(\omega_{n},\bmp)} \nonumber \\
& & \times\{ {\bf \Delta}(\bmp)(p^{2}_{0}\mp\bmp^{2}-m^{2}-\mu^{2}-{\bf \Delta}(\bmp)\cdot{\bf \Delta}^{*}(\bmp))\pm 2\bmp({\bf \Delta}(\bmp)\cdot\bmp) \}.
\end{eqnarray}
Here the upper sign corresponds to the case of the axial vector while the lower sign corresponds to the case of the axial-vector-like components of the 2-rank antisymmetric tensor. We take the expansions of the interaction and mean fields as discussed previously. Throughout our study, we take only $l=1$ p-wave interaction into account. For the mean fields, we treat $j=0, 1, 2$ states. We can consider coupled states with different $j, m_{j}, \lambda$, but hereafter we only treat definite values of $j, m_{j}, \lambda$, because of simplicity. In the general form, 
\begin{eqnarray}
\Delta^{(\lambda)}_{jm_{j}}\psi_{jm_{j}\lambda}(\hat{\bmk}) &=& \int\frac{d^{3}\bmp}{(2\pi)^{3}}\sum_{n}\frac{1}{\beta}4\pi g_{1}\sum_{m}Y_{1m}(\hat{\bmk})Y^{*}_{1m}(\hat{\bmp})\frac{1}{D(\omega_{n},\bmp)} \nonumber \\
& & \times\{ \Delta_{jm_{j}}^{(\lambda)}\psi_{jm_{j}\lambda}(\hat{\bmp})(p^{2}_{0}\mp\bmp^{2}-m^{2}-\mu^{2}-|\Delta^{(\lambda)}_{jm_{j}}|^{2}\psi^{*}_{jm_{j}\lambda}(\hat{\bmp})\cdot\psi_{jm_{j}\lambda}(\hat{\bmp})) \nonumber \\
& & \pm\bmp(\Delta^{(\lambda)}_{jm_{j}}(\psi_{jm_{j}\lambda}(\hat{\bmp})\cdot\bmp))\},   
\end{eqnarray}
we multiply both sides with $\psi^{*}_{jm_{j}\lambda}(\hat{\bmk})$ from the left, and perform angular integration with respect to the spherical angle of $\hat{\bmk}$. After a lengthy manipulation, we obtain 18 gap equations, and all of them have the forms of the following 2-type gap equations, either Type-(i) or Type-(ii):

Type-(i)
\begin{eqnarray}
1 &=& ag_{1}\int\frac{d^{3}\bmp}{(2\pi)^{3}}c(\theta)\biggl\{\biggl(1-\frac{\bmp^{2}}{\sqrt{(\bmp^{2}+m^{2})\mu^{2}+b|\Delta^{(\lambda)}_{jm_{j}}|^{2}\bmp^{2}c(\theta)}}\biggr)\frac{1}{2E_{+}}\tanh\frac{\beta}{2}E_{+} \nonumber \\
& & +\biggl(1+\frac{\bmp^{2}}{\sqrt{(\bmp^{2}+m^{2})\mu^{2}+b|\Delta^{(\lambda)}_{jm_{j}}|^{2}\bmp^{2}c(\theta)}}\biggr)\frac{1}{2E_{-}}\tanh\frac{\beta}{2}E_{-}\biggr\}, \\ 
E_{\pm} &=& \sqrt{\bmp^{2}+m^{2}+\mu^{2}+b|\Delta^{(\lambda)}_{jm_{j}}|^{2}c(\theta)\mp2\sqrt{(\bmp^{2}+m^{2})\mu^{2}+b|\Delta^{(\lambda)}_{jm_{j}}|^{2}\bmp^{2}c(\theta)}}. 
\end{eqnarray}

Type-(ii) 
\begin{eqnarray}
1 &=& ag_{1}\int\frac{d^{3}\bmp}{(2\pi)^{3}}c(\theta)\biggl(\frac{1}{2E_{+}}\tanh\frac{\beta}{2}E_{+}+\frac{1}{2E_{-}}\tanh\frac{\beta}{2}E_{-}\biggr), \\
E_{\pm} &=& \sqrt{(\sqrt{\bmp^{2}+m^{2}}\mp\mu)^{2}+b|\Delta^{(\lambda)}_{jm_{j}}|^{2}c(\theta)}. 
\end{eqnarray}
In the above equations, $a$, $b$ and $c(\theta)$ depend on specific $\psi_{jm_{j}\lambda}$. All the 18 gap equations are listed in table I. In this table, we also list the node structure of the gap functions.  

About $\psi_{000}$ cases, there is no angular dependence both in the eigenvalues and gap equations. The condensation is for $j=l+s=0$ pairs. $\psi_{000}$ state of ${\bf \Delta}^{T}_{(A)}$ corresponds to the BW ( Balian-Werthamer ) state in our theory~[9,35]. The gap is isotropic like the nonrelativistic BW state. The structure of this gap equation is the same as the $\Delta^{S}$ given above ( Eqs. (104) and (105) ). $\psi^{(-)}_{101}$ state of ${\bf \Delta}^{A}$ can be regard as the ABM ( Anderson-Brinkman-Morel ) state in our theory~[8,35]. In this case, the spin vector is orthogonal to the vector of orbital motion. The angular dependence of this gap in the gap equation is the same as $Y_{1\pm1}$, like the nonrelativistic ABM. It has nodes at two points $\theta=0, \pi$. There is no state of $j=1$ which has a line node structure. It is interesting that, in spite of the spin triplet pairing, the structures of the gap nodes of $j=2$ states are like that of d-wave, $Y_{2m}$. This feature arises from our relativistic treatment, i.e. the use of the helicity state. 

In the spin-triplet pairing states discussed above, the quasiparticle dispersion and the gap equation take simple forms like the case of $\Delta^{S}$ except the angular dependences of the gaps ( Type-(ii) ), for the axial vector when the orientation of the mean fields is perpendicular to $\hat{\bmk}$ ( helicity $\pm 1$, the ``transversal'' ), and for the 2-rank antisymmetric tensor when the direction of the mean fields is parallel with $\hat{\bmk}$ ( helicity $0$, the ``longitudinal''). Other cases, the eigenvalue and the gap equation have complicated structure in Type-(i), and they are similar to that of the $\Delta^{V}_{0}$. These are clear from the form of the functions, (99), (101) and (112). Thus in the relativistic theory, the difference in the helicity ( it gives the orientation of the mean field vector ) gives rise to a quite different form for both the dispersion and gap equation. It might be interesting if the relative orientation between the mean field and $\bmk$ gives large effects to the solutions of the gap equations, thermal properties or response of the system to the external fields. These gap equations will be studied numerically in part II of this paper.

\section{Summary}
\label{sec:con}

In this paper, we have performed the investigation of BCS and generalized BCS superconductivity in relativistic quantum field theory. We have introduced the Gor'kov equation, and given the group theoretical considerations about the superconducting mean fields in complete manner. Especially to treat the spin triplet Cooper pairs, we have investigated the generalized BCS formalism in our theory. We have solved the Gor'kov equations completely under the assumption to have the specific types of the mean fields with unitary condition.  We have constructed the gap equations in the various states and discussed their details. Throughout this paper, we have discussed various characteristic features of the theory. 

As a next work, we will present the results of solving the gap equations numerically. We will also give some results of the thermodynamics, or response to the external fields. Preparation for the presentation of these results is now in progress, and will be published as part II of this study.

\acknowledgements

The author would like to thank Profs. H. Akai, K. Higashijima, Y. Hosotani, K. Ishikawa, Y. Nambu, H. Toki and K. Yamaguchi for many helpful discussions.

\begin{table}
\caption{The coefficients in the gap equations of the ${\bf \Delta}^{A}$ and ${\bf \Delta}^{T}_{(A)}$ pairings. (i) and (ii) correspond to the use of the gap equations of Type-(i) and Type-(ii), in (114,115) and (116,117), respectively. $a$, $b$ and $c(\theta)$ in this table are the coefficients in these equations. }
\begin{tabular}{c|cccccc}
&\multicolumn{2}{c}{type}&&&&\\
\cline{2-3}
state & ${\bf \Delta}^{A}$ & ${\bf \Delta}^{T}_{(A)}$ & $a$ & $b$ & $c(\theta)$& node \\
\tableline
$\psi_{000}$ & (i) & (ii) & $\frac{1}{2}$ & $\frac{1}{4\pi}$ & $1$ & no node \\
$\psi_{200}$ & (i) & (ii) & $\frac{1}{4}$ & $\frac{5}{16\pi}$ & $(3\cos^{2}\theta-1)^{2}$ & $\cos\theta=\pm\frac{1}{\sqrt{3}}$ \\
$\psi_{210}$ & (i) & (ii) & $\frac{3}{2}$ & $\frac{15}{8\pi}$ & $\sin^{2}\theta\cos^{2}\theta$  & $\theta=0,\frac{\pi}{2},\pi$ \\
$\psi_{220}$ & (i) & (ii) & $\frac{3}{8}$ & $\frac{15}{32\pi}$ & $\sin^{4}\theta$ & $\theta=0,\pi$\\
$\psi^{(-)}_{101}$ & (ii) & (i) & $\frac{3}{4}$ & $\frac{3}{8\pi}$ & $\sin^{2}\theta$ & $\theta=0,\pi$\\
$\frac{1}{\sqrt{2}}(\psi^{(-)}_{111}-\psi^{(-)}_{1-11})$ & (ii) & (i) & $\frac{3}{8}$ & $\frac{3}{16\pi}$ & $\cos^{2}\theta+1$ & no node\\
$\psi^{(+)}_{201}$ & (ii) & (i) & $\frac{9}{4}$ & $\frac{15}{8\pi}$ & $\sin^{2}\theta\cos^{2}\theta$ & $\theta=0,\frac{\pi}{2},\pi$ \\
$\frac{1}{\sqrt{2}}(\psi^{(+)}_{211}-\psi^{(+)}_{2-11})$ & (ii) & (i) & $\frac{3}{8}$ & $\frac{5}{16\pi}$ & $4\cos^{4}\theta-3\cos^{2}\theta+1$ & no node\\
$\frac{1}{\sqrt{2}}(\psi^{(+)}_{221}+\psi^{(+)}_{2-21})$ & (ii) & (i) & $\frac{3}{8}$ & $\frac{5}{16\pi}$ & $1-\cos^{4}\theta$ & $\theta=0,\pi$ \\
\end{tabular}
\end{table}

\end{document}